\documentclass[twocolumn,showpacs,preprintnumbers]{revtex4}
\usepackage{graphicx}
\usepackage{dcolumn}
\usepackage{bm}

\draft

\begin{document}

\title{Micro-Raman evidence for topological charge order across the superconducting dome of La$_{2-x}$Sr$_x$CuO$_{4}$}
\author{D. Lampakis, E. Liarokapis}
\address{Department of Physics, National Technical University,\\
GR-15780 Athens, Greece}
\author{C. Panagopoulos}
\address{Cavendish Laboratory and IRC in Superconductivity, University of Cambridge,\\
Madingley Road, CB3 0HE, United Kingdom}
\date{\today}

\begin{abstract}

The doping dependence of the micro-Raman spectra of high quality
La$_{\rm 2-x}$Sr$_{\rm x}$CuO$_{\rm 4}$ microcrystals with
x=0.0-0.45 has been investigated in the region 10-300 K. The
phonon at $\sim $100 cm$^{-1}$ of the orthorhombic phase shows a
classical soft mode behavior up to x=0.17, supporting its
correlation with the LTO to HTT transition. In the xx/yy
polarization spectra of the superconducting concentrations new
modes at $\sim $150 and 370 cm$^{{\rm -} {\rm 1}}$ related to the
symmetry breaking and a broad band at $\sim $280 cm$^{{\rm -} {\rm
1}}$ are attributed to the charge ordering at temperatures well
above T$_{\rm c}$. At all temperatures studied a correlation has
been found between the doping dependence of the transition
temperature and the intensity of the bands at $\sim $150,
$\sim$280, $\sim$370 cm$^{{\rm -} {\rm 1}}$, the intensity of the
La/Sr mode, and the asymmetry of the apex phonon.
\end{abstract}

] \narrowtext
\maketitle

\subsection*{INTRODUCTION}

It is known that the La$_{{\rm 2}{\rm -} {\rm x}}$Sr$_{{\rm
x}}$CuO$_{{\rm 4}}$ (LSCO) superconductor shows a variety of
phases with doping and temperature \cite{Keimer}. One of the main
conclusions from the Raman investigations of the vibrational modes
of La$_{\rm 2}$CuO$_{{\rm 4}}$ has been the existence of a soft
mode at $\sim $100 cm$^{{\rm -} {\rm 1}}$ (12 meV) of A$_{{\rm
g}}$ symmetry, attributed to tilting vibrations of the CuO$_{{\rm
6}}$ octahedra about the diagonal (110) axis
\cite{Sugai,Weber,Burns}. This mode has been correlated with the
observed High Temperature Tetragonal (HTT) to Low Temperature
Orthorhombic (LTO) structural phase transition of LSCO
\cite{Birgeneau} and has been studied for Sr doped samples, up to
x=0.07 \cite{Sugai,Weber,Lampakis}. As the structural critical
temperature decreases with increasing Sr content \cite{Keimer},
the soft mode can only be observed at LT and it has not been
studied for x$>$0.07. Here we examine the behavior of the soft
mode for even higher Sr concentrations and, particularly, for x
values near optimum.

The two strong modes of this compound are of A$_{{\rm g}}$
symmetry and have been attributed to the vibrations of the La/Sr
and the apex oxygen (O$_{{\rm z}}$) atoms along the c axis. Their
dependance on temperature has been studied for a wide range of Sr
doping. Our studies of these materials at room temperature (RT)
have shown that, the asymmetry of the apex phonon and the relative
intensity of the La/Sr phonon over the apex mode show a maximum
for the Sr concentrations at which the compound appears with the
maximum transition temperature to superconductivity
\cite{Lampakis}. It was concluded that, since this effect cannot
be assigned to any apparent structural changes of the system, it
might originate from a charge redistribution, which occurs at
temperatures well above T$_c$ and affects the electronic states
and the polarizabilities of the La/Sr and apex atoms at some
critical Sr concentration. For this reason, it was interesting to
examine how this correlation of some phonon characteristics with
the transition temperature evolves at low temperatures.

Furthermore, we found that, with increasing x two symmetry
forbidden broad modes at $\sim $150 and $\sim$370 cm$^{{\rm -}
{\rm 1}}$ appear in the xx/yy polarization Raman spectra, despite
the phase transition from the orthorhombic to the more symmetric
tetragonal phase \cite{Sugai,Weber}. In the low temperature
spectra a broad band at $\sim $280 cm$^{{\rm -} {\rm 1}}$ has been
also observed \cite{Sugai}. A common feature of these peaks is
that their intensity shows a maximum for the optimally doped
samples \cite{Lampakis}. In this work we have investigated the
temperature dependence of the Raman spectra for various Sr
contents, and compare with previously reported room temperature
data.

\subsection*{EXPERIMENTAL SETUP}

Individual microcrystallites from a selected series of high
quality La$_{{\rm 2}{\rm -} {\rm x}}$Sr$_{{\rm x}}$CuO$_{{\rm 4}}$
polycrystalline compounds, with Sr doping in the range 0.00$ \le
$x$ \le $0.45 have been studied in the 10-300 K temperature
region. The carefully prepared samples have been characterized as
described elsewhere \cite{Lampakis}. The Raman spectra were
obtained in the approximate y(zz)$\bar {y}$ and y(xx)$\bar {y}$
(or x(yy)$\bar {x}$) scattering configurations or a mixture of
them (as the x and y axis could not be discriminated in the
twinned samples) at nominal temperatures from 10K to RT with a
Jobin-Yvon T64000 triple spectrometer, equipped with a liquid
nitrogen cooled CCD and a microscope ($\times $100 magnification).
Low temperatures were achieved by using an open cycle Oxford
cryostat operating with liquid nitrogen or with liquid helium. The
514.5 nm of an Ar$^{{\rm +} }$ laser was used for excitation at a
power level of $ \le $0.1 mW/$\mu$m$^{{\rm 2}}$. The local sample
heating due to the laser beam was estimated to be less than 10K.
Accumulation times were of the order of 4-5 hours and usually two
microcrystallites have been studied for each Strontium
concentration (i.e. x= 0.00, 0.03, 0.05, 0.07, 0.092, 0.125, 0.17,
0.20, and 0.24).

\subsection*{RESULTS}

Typical spectra for selected doping and temperatures in the
y(zz)$\bar {y}$ and y(xx)$\bar {y}$ (or x(yy)$\bar {x}$)
polarization, are shown in Figures 1 and 2, respectively. In all
cases, the two strong La/Sr and apex oxygen phonons of A$_{{\rm
g}}$ symmetry appear. According to the La$_{{\rm 2}{\rm - }{\rm
x}}$Sr$_{{\rm x}}$CuO$_{{\rm 4}}$ phase diagram, the structural
critical transition temperature, T$_{{\rm o}{\rm t}}$, for the
insulating La$_{{\rm 1}{\rm .}{\rm 9}{\rm 7}}$Sr$_{{\rm 0}{\rm
.}{\rm 0}{\rm 3}}$CuO$_{{\rm 4}}$ sample is $\sim $560K
\cite{Keimer}. Thus, this compound remains orthorhombic throughout
the temperature range studied and the soft mode appears in the low
energy part of the zz polarization spectra for all temperatures
(Fig.1a). The mode at $\sim $270 cm$^{{\rm -} {\rm 1}}$ attributed
to the vibrations of the plane oxygens along the c axis is
observed in all zz spectra of Fig.1a. This phonon is also
correlated with the HTT$\rightarrow$LTO structural phase
transition since the vibrations of the plane oxygens are not Raman
active in the tetragonal phase \cite{Sugai,Weber,Bazhenov}.
Concerning the xx/yy polarization spectra for the x=0.03 sample
(Fig.2a), new bands at $\sim $150 and 370 cm$^{{\rm -} {\rm 1}}$
are observed for temperatures below $\sim $100 K. These broad
peaks are weakly observed at the same low temperatures in the zz
polarization spectra (Fig.1a). With increasing doping these two
bands appear at much higher temperatures, even at RT in the
spectra of the superconducting materials (tetragonal phase,
Fig.2b-2e). Therefore, their appearance cannot be related with the
HTT$\rightarrow$LTO structural phase transition. Besides, a weak
mode at $\sim $320 cm$^{{\rm -} {\rm 1}}$ of A$_{{\rm 1}{\rm g}}$
symmetry \cite{Weber} is present in the xx/yy spectra with energy
independent of temperature (Fig.2a). A broad continuum around 280
cm$^{-1}$ is also present at low temperatures and for intermediate
doping in the same polarization spectra (Fig.2c-2d). Finally, the
mode at $\sim $100 cm$^{{\rm -} {\rm 1}}$ rapidly softening with
increasing temperature (Fig.2a) is apparently an escape from the
zz polarization.

In the zz polarization spectra of the underdoped superconductor
La$_{{\rm 1}{\rm .}{\rm 9}{\rm 0}{\rm 8}}$Sr $_{{\rm 0}{\rm .}{\rm
0}{\rm 9}{\rm 2}}$CuO$_{{\rm 4}}$ with T$_{{\rm o}{\rm t}}\sim
$300K, the modes at $\sim $100 and 270 cm$^{{\rm - }{\rm 1}}$ are
observed only below 250 K (Fig.1b). For T$>$250 K they are not
detectable due to the small tilting of the octahedra, in agreement
with their relation to the orthorhombic deformation of the cell.
The weak peaks at $\sim $150 cm$^{{\rm -} {\rm 1}}$ and $\sim $365
cm$^{{\rm -} {\rm 1}}$ must be the same modes with the xx/yy
polarization spectra (see Fig.2b). The very broad band at $\sim
$280 cm$^{{\rm -} {\rm 1}}$ observed in xx/yy polarization at low
temperatures with increasing intensity should not be confused with
the narrow peak at $\sim $270 cm$^{{\rm -} {\rm 1}}$ of the plane
oxygens that appears in the zz polarization spectra. The intensity
increment of the three bands at $\sim$150, $\sim$280, and
$\sim$370 cm$^{{\rm -} {\rm 1}}$ with decreasing temperature
becomes more obvious in the samples with x=0.125 and 0.17
(Fig.2c,d). In La$_{{\rm 1}{\rm .}{\rm 7}{\rm 6}}$Sr$_{{\rm 0}{\rm
.}{\rm 2}{\rm 4}}$CuO$_{{\rm 4}}$ the intensity of the three broad
bands appears substantially reduced (Fig.2e). This indicates a
doping dependence of these bands and, possibly small lattice
distortions induced by the variation of the Sr content. The
dependence of these bands on temperature and Sr concentration is
examined below.

For the x=0.125 compound, some traces of the soft mode can be
detected in the zz spectrum at 250 K with increasing intensity at
lower temperatures (Fig.1c). Concerning the x=0.17 sample the soft
mode start appearing in the zz spectra at even lower temperatures
(Fig.1d), in agreement with the phase diagram, where no tilting is
expected in the tetragonal phase above 150 K \cite{Keimer}. The
soft mode has not been detected in the zz spectra for any
temperature in the overdoped superconductor La$_{{\rm 1}{\rm
.}{\rm 7}{\rm 6}}$Sr$_{{\rm 0}{\rm .}{\rm 2}{\rm 4}}$CuO$_{{\rm
4}}$ (Fig.1e) and for x=0.20 as expected, since the compound
remains tetragonal at any temperature \cite{Keimer}.

The soft mode behavior of the low energy A$_{{\rm g}}$ mode at
$\sim $100 cm$^{{\rm -} {\rm 1}}$ has already been shown for low
Sr concentrations, up to x=0.07 \cite{Sugai,Weber,Lampakis}. In
Figure 3a the temperature dependence of the energy of this mode
for selected concentrations is presented. The values of the energy
were determined by fitting the low energy part of the Raman
spectra with Lorentzian curves. In Fig.3a, the continuous lines
are best fit to the data with curves $\omega_{soft} \propto \rm
(T_{ot} -T)^{1/2}$, which assume a 2nd order phase transition. For
most concentrations the zero phonon energy occurs close to the
structural critical temperature of each concentration. But, for
the x=0.17 sample, the soft mode energy tends to zero at
temperature fairly larger than T$_{{\rm o}{\rm t}}$. This
discrepancy may originate from the fact that the soft mode peak
for this sample is very weak for all temperatures studied, and,
therefore, the accurate determination of the energy was very
difficult. Figure 3b shows the temperature dependence of the
linewidth of the soft mode, for the x=0.092-0.17 samples. In each
case, on approaching the T$_{{\rm o}{\rm t}}$ the width increases
as in a second order phase transition. Also, it should be noticed
that, the soft mode is very weak for temperatures above 200 K, for
the x=0.092, 0.125, and 0.17 samples (Fig.1b, c, d), and,
therefore, the fitting results above this temperature are not
taken into account in graphs of Fig.3. The above results support
the correlation of this mode with the structural phase transition
HTT$\rightarrow$LTO. As seen in Fig.4, the energy of the soft mode
decreases with increasing Sr content while its width increases as
expected from the reduced tilting of the octahedra with increasing
doping.

As mentioned, the mode at $\sim $270 cm$^{{\rm -} {\rm 1}}$
observed in the zz scattering polarization is attributed to the
vibrations of the plane oxygens along the c axis of A$_{{\rm g}}$
symmetry. It is correlated with the structural phase transition,
since it appears only in the spectra of the orthorhombic phase of
the compound, just like the soft mode. In particular, for the
insulating samples (Fig.1a for x=0.03) this mode appears in the
whole temperature range 10-300 K, while, for the doped x=0.092 and
0.125 samples it can only be seen at temperatures below 250 K. For
the optimally doped x=0.17 sample, the appearance of this mode at
temperatures above 150 K (which is the nominal T$_{{\rm ot}}$ for
this compound), just like the soft mode, supports the assumption
of the existence of tilted octahedra at this concentration and,
thus, the T$_{{\rm ot}}$ for this sample should be considered
$\sim $200 K. For the overdoped samples this mode has not been
detected in all temperatures studied. Although the weakness of
this mode and the existence of the strong La/Sr phonon close to it
make difficult the accurate determination of the mode energy and
width, it can be clearly seen that there is an almost linear
increase in energy with decreasing temperature of the order $\sim
$5-6 cm$^{{\rm -} {\rm 1}}$ (Fig.1). This result is in
disagreement with previously reported data where no energy shift
with temperature has been observed for this mode
\cite{Weber,Sugai}. In addition, it should be clarified that,
although the appearance of this mode is correlated with the
LTO$\rightarrow$HTT transition, it does not show a classical soft
mode behavior, i.e., its energy does not tend to zero near
T$_{{\rm o}{\rm t}}$ and its width does not show a significant
broadening, similar to the broadening of the mode at $\sim $100
cm$^{{\rm -} {\rm 1}}$.

In Figure 5a the variation of the energy of the La/Sr mode as a
function of temperature for the x=0.092-0.24 superconducting
samples is presented. As seen, the phonon energy decreases almost
linearly with temperature for each Sr content by approximately 3-4
cm$^{{\rm -} {\rm 1}}$ in the range 10-300 K. A similar behavior
has also been detected for the insulating x=0.0, x=0.03 (also
shown in Fig.5a) and underdoped superconducting (x=0.05, 0.07)
compounds \cite{Lampakis}. On the contrary, the apex energy
remains practically constant with temperature for all  compounds
in the x=0.0-0.17 range (Fig.5b and Ref.\cite{Lampakis}). Instead,
there is a decrease of approximately 3 cm$^{{\rm -} {\rm 1}}$,
with increasing temperature for the overdoped (x=0.24) compound.
It should be marked that for the optimally doped compounds
(x=0.125 and 0.17), the material undergoes a structural phase
transition HTT$\rightarrow$LTO phase in the temperature range
studied. This transition is correlated with the tilting of the
CuO$_{{\rm 6}}$ octahedra about the (110) axis and apparently it
does not affect the energy of the apex oxygen.

Figure 6 shows the linewidths of the La/Sr and apex oxygen modes
with varying temperature for several doping levels. For all Sr
concentrations the linewidth of both modes shows a typical
anharmonic behavior increasing almost linearly with temperature
above 70 K. The total width increment is about 8 cm$^{{\rm -} {\rm
1}}$, for both phonons in all concentrations. The relative
intensity of the two modes as a function of temperature is
presented in Fig.7 for the same concentrations. In a first
approximation the relative intensity increases with temperature
for all concentrations studied. But it appears that in the
temperature range $\sim$150-250 K there are some profound
deviations from linearity for the samples around optimal doping.
The maximum deviation is for the x=0.125 compound, it is reduced
for x=0.17 and it seems gradually to disappear for x=0.092, and
x=0.24. \textit{Since no structural modifications are expected at
low temperatures for these compounds, this change perhaps involves
other mechanisms and it is probably related with the plateau in
the doping dependence of T$_c$ and the charge ordering}
\cite{Bianconi}. It is also clear that the correlation observed at
RT between this relative intensity and the variation of T$_c$ with
doping \cite{Lampakis} remains at low temperatures.

Figure 8 shows the temperature dependence of the average energy
and total apparent width of the wide band at $\sim $150 cm$^{{\rm
- }{\rm 1}}$. At room temperature there is a band energy
difference among the various concentrations, which diminishes at
low temperatures. Besides, the temperature dependence of the band
energy depends on the amount of doping. The bandwidth is almost
independent of temperature for x$\geq$0.20, while for lower doping
levels it increases with decreasing temperature (Fig.8). One can
observe in Figure 2 that this band for certain concentrations and
below some temperature splits in two or three modes. Therefore,
the modifications shown in Fig.8 actually represent changes in the
relative intensity and the width of those modes.

Figure 9 shows the temperature dependence of the energy and width
of the symmetry forbidden mode at $\sim $370 cm$^{{\rm - }{\rm
1}}$. For the x=0.092-0.24 compounds the energy and the width of
the mode seems to be independent of doping and temperature. In the
insulating sample (x=0.03) a mode of similar energy appears in the
xx/yy spectra for temperatures below 100 K (Fig.2a). The energy of
this mode depends strongly on temperature decreasing by $\sim $8
cm$^{{\rm - }{\rm 1}}$ from T=90 to 10 K. Its linewidth is
initially narrow, but with decreasing temperature it increases
reaching at 10 K almost the same value as for the x=0.092-0.24
samples (Fig.9b).

The broad band at $\sim $280 cm$^{{\rm -} {\rm 1}}$ appears mainly
in the low temperature xx/yy spectra. At room temperature, traces
of this band have been detected only for the x=0.17 sample
(Fig.2d). For x=0.092 and 0.125 the band is observed for
temperatures below 250 K (Figs.2b and 2c), while, for the x=0.03
sample it seems not to be present to the lowest temperatures
studied (Fig.2a). Similar to the $\sim $150 and 370 cm$^{{\rm -}
{\rm 1}}$ broad peaks, the appearance of this band cannot be
correlated with Raman active modes or the HTT$\rightarrow$LTO
phase transition. For x=0.125 the transition temperature T$_{ot}$
is nearly the same with the temperature at which the $\sim $280
cm$^{{\rm -} {\rm 1}{\rm} }$ band appears, but this is not the
case for the x=0.092 and x=0.17 samples. Besides, the x=0.03,
which is orthorhombic even at room temperature does not show this
broad band. For the x=0.092-0.17 compounds, the energy of this
band does not vary with temperature within experimental error
(Fig.10a), while its width appears to be very large ($\sim $100
cm$^{{\rm -} {\rm 1}}$) and almost independent of temperature for
all samples (Fig.10b).

\subsection*{DISCUSSION}

The apex phonon energy shown in Fig.5 (and from
Ref.\cite{Lampakis}) remains practically unaffected with the
variation of temperature for the x=0.03-0.17 samples and, only,
for the overdoped x=0.24 sample seems to decrease with increasing
temperature ($\sim $3 cm$^{{\rm -} {\rm 1}}$). This result for the
x=0.03-0.17 compounds is in good agreement with the structural
measurements, which have shown that the Cu-O$_{{\rm z}}$ bond is
practically independent of temperature or varies by less than
0.2\% for all doping levels \cite{Radaelli}. For the overdoped
x=0.24 compound the disagreement for the overdoped region has also
been observed at RT as a function of doping \cite{Lampakis}, where
the Cu-O$_{{\rm z}}$ bond decreases with increasing Sr
concentration together with the energy of the apex oxygen mode.

Bianconi et al. \cite{Bianconi} in an  EXAFS study has evaluated
the Cu-O$_{{\rm x}{\rm y}}$ and Cu-O$_{{\rm z}}$ bond lengths of
the optimally doped La$_{{\rm 1}{\rm .}{\rm 8}{\rm 5}}$Sr$_{{\rm
0}{\rm .}{\rm 1}{\rm 5}}$CuO$_{{\rm 4}}$ compound. They have shown
that there are two Cu-O$_{{\rm z}}$ bond lengths that differ by
0.08 to 0.14{\AA}, depending on the temperature of the material.
Also, two different Cu-O$_{{\rm x}{\rm y}}$ bonds have been
observed for this material, but only for temperatures below 100 K
(i.e. in the orthorhombic phase), whereas the tilting angle of the
octahedra was found to be larger. From the above it was concluded
that in the La$_{{\rm 1}{\rm .}{\rm 8}{\rm 5}}$Sr$_{{\rm 0}{\rm
.}{\rm 1}{\rm 5}}$CuO$_{{\rm 4}}$ structure apart of the normal
(undistorted) there are also distorted octahedra. These local
lattice distortions have been attributed to the existence of two
types of stripes, the U-type, which is due to the undistorted
octahedra, and the D-type which is due to the distorted ones
\cite{Bianconi}.

The existence of a shorter Cu-O$_{{\rm z}}$ bond would imply
another weak phonon peak in the Raman spectra at the high energy
side of the band due to the vibrations of the apex oxygen atoms
along the c-axis (A$_{{\rm g}}$ symmetry). A weak peak could be
related with the non-variation of the apex phonon energy with
temperature for the compounds with x$ \le $0.17. The decrease of
the energy of this mode with temperature for the x=0.24 sample
would imply the absence of the shorter Cu-O$_{{\rm z}}$ bond in
the overdoped region. Also, the asymmetry of the A$_{{\rm g}}$
apex phonon may originate from doping induced local lattice
distortions and the second phonon peak. A correlation has already
been established at room temperature between the doping dependence
of the asymmetry of this phonon (as observed in the
zz-polarization) and the relative dependence of T$_{{\rm c}}$
\cite{Lampakis}. As can be seen in Fig.11 a similar correlation
between the apex phonon asymmetry and T$_c$ exists at the lowest
temperatures studied. For the xx-polarization spectra (Fig.2) the
asymmetry of the apex mode is even more pronounced. For x=0.03 and
0.24 there is no asymmetry for all temperatures (Fig.2a,2e), while
for the other concentrations the asymmetry of the apex phonon is
quite clear. For x=0.125 the apex phonon seems to consist at least
of another peak at the high energy side. This critical
concentration corresponds to the plateau in the T$_{{\rm c}}$.

It is known that, at room temperature, the relative intensity of
the two strong A$_{{\rm g}}$ phonons varies with doping in a
similar way as T$_{{\rm c}}$ \cite{Lampakis}. According to the low
temperature data, the relative intensity La/apex as a function of
Sr content reaches a maximum value in the optimally doped region
for all temperatures studied (Fig.12). The substitution of
Sr$^{{\rm +} {\rm 2}}$ for La$^{{\rm +} {\rm 3}}$ in LSCO adds
holes to the compound from the La/Sr sites, with the
superconductivity appearing when the carrier concentration at the
CuO$_{{\rm 2}}$ planes exceeds a limit. The small intensity of the
La/Sr A$_g$ phonon for x$\leq$0.03 in Fig.12 can be attributed to
the local screening of the light due to the high concentration of
carriers at these sites. By increasing the amount of doping, the
holes at the La/Sr sites are expected to move to the CuO$_2$
planes reducing the screening effect and, consequently, increasing
the La/Sr phonon intensity. With the further increase of doping
above optimal the La/apex relative intensity seems to decrease
again, but at much less rate than the initial increase for
x=0.0-0.03 (Fig.12). It is possible that after the optimal doping
the additional carriers can no longer move preferentially to the
CuO$_{{\rm 2}}$ planes screening equally all phonons.

In the xx/yy scattering polarization and for doping concentrations
that the compound is in the tetragonal phase and superconducting,
new symmetry-forbidden modes appear in the Raman spectra (Fig.2).
The space group of the tetragonal structure (D$_{{\rm 4}{\rm
h}}^{{\rm 1}{\rm 7}}$) is symorphic with inversion center the Cu
site. Since no symmetry lowering has been observed in the
crystallographic studies of these materials, the appearance of new
modes denotes a local symmetry breaking of the tetragonal
structure. This symmetry breaking can be induced by the partial
substitution of Sr for La, but then it should be present in any
concentration of Sr continuously increasing with the amount of
doping, which does not agree with our Raman data (absence of the
$\sim $370 cm$^{{\rm -} {\rm 1}}$ mode and substantial reduction
in the intensity of the $\sim $150 cm$^{{\rm -} {\rm 1}}$ for
x=0.24, Fig.2e).

From their energies, the bands at $\sim $150 cm$^{{\rm -} {\rm
1}}$ and $\sim $370 cm$^{{\rm -} {\rm 1}}$ could be assigned to
IR-active phonons. In the E//ab polarization IR spectra of the
tetragonal phase a doubly degenerate E$_{{\rm u}}$ mode exists at
$\sim $140 cm$^{{\rm -} {\rm 1}{\rm} }$ \cite{Sugai,Bazhenov},
attributed to the vibrations of the apex oxygen atoms parallel to
the ab-plane. In the orthorhombic phase the degeneracy is removed
and for x=0.0 the E$_{{\rm u}}$ mode splits to one at $\sim $113
cm$^{{\rm -} {\rm 1}}$ of B$_{{\rm 2}{\rm u}}$ symmetry
(vibrations of the O$_{{\rm ap}}$ atoms along the b-axis) and
another at 140 cm$^{{\rm -} {\rm 1}}$ of B$_{{\rm 3}{\rm u}}$
symmetry (vibrations of the O$_{{\rm ap}}$ atoms along the a-axis)
\cite{Bazhenov}. The XRD measurements have shown \cite{Radaelli},
that the b-axis increases substantially in the orthorhombic phase.
Therefore, the above splitting depends on the amount of
orthorhombic distortion reducing to zero in the tetragonal phase.
In the Raman spectra of the orthorhombic phase the $\sim $150
cm$^{{\rm -} {\rm 1}}$ band appears at low temperatures as a
double or triple peak. The new mode at $\sim$144 cm$^{{\rm -} {\rm
1}}$ becomes very narrow and the other one at $\sim$125 cm$^{{\rm
-} {\rm 1}}$very broad at low temperatures and doping levels
x=0.092-0.17 (Fig.2b, c, and d). Besides, there are traces of a
peak at $\sim$170 cm$^{{\rm -} {\rm 1}}$, which could correspond
to another IR mode \cite{Padilla} (Fig.2a). The apparent increment
of the total width of the $\sim $150 cm$^{{\rm -} {\rm 1}}$ band
with decreasing temperature for the x=0.092-0.17 samples (Fig.8)
obviously reflects modifications in the width and intensity of the
split modes.

Our data in Fig.2 indicate that the splitting of the $\sim$150
cm$^{{\rm -} {\rm 1}}$ band occurs in the range 200-250 K for
x=0.092, 150-200 K for x=0.125, and 150-200 K for x=0.17. In the
x=0.20 and 0.24 samples, the structural measurements have shown a
tetragonal structure at any temperature and the $\sim $125
cm$^{{\rm -} {\rm 1}}$ wide mode does not appear to the lowest
temperatures (Fig.2e). In the x=0.03 compound the $\sim $150
cm$^{{\rm -} {\rm 1}}$ band is hardly observed (Fig.2a), while for
x=0.07 the wide mode at $\sim $125 cm$^{{\rm -} {\rm 1}}$ is
detected above 125 K. Once the mode at $\sim$144 cm$^{{\rm -} {\rm
1}}$ (that can be associated with the IR mode of symmetry
vibrations of the O$_{{\rm z}}$ atoms along the a-axis) appears it
does not vary in energy with decreasing temperature. On the other
hand, the mode at $\sim$125 cm$^{{\rm -} {\rm 1}}$ (associated
with the vibrations of the O$_{{\rm z}}$ atoms along the b-axis)
becomes broader and softens at decreasing temperatures. This could
be due to the appearance of another IR mode at lower energy
\cite{Padilla}. The Raman spectra with the many new lines that
appear at low temperatures are very reminiscent of the formation
of stripes in related compounds \cite{stripes} and fits very well
with the findings of Bianconi et al \cite{Bianconi}. The charge
ordering is expected to give rise to the formation of a
superlatice, lowering the crystal symmetry and bringing edge
phonons to the $\Gamma$ point. Neutron scattering data on
LaCuO$_4$ have found a mode of A$_{\rm g}$ symmetry at the X point
from the La atoms of similar energy with the 150 cm$^{{\rm -} {\rm
1}}$ band \cite{Pintschovius}. Therefore, the observed multiple
modes could also be due to such folded phonons unambiguously
defining the formation of stripes.

Concerning the possible assignment of the $\sim $370 cm$^{{\rm -}
{\rm 1}}$ band to a phonon, it can be seen that in the IR spectra
of the tetragonal phase a q$\simeq$0 mode of E$_{{\rm u}}$
symmetry appears at a similar energy ($\sim $360 cm$^{{\rm -} {\rm
1}}$) \cite{Sugai,Bazhenov}. This mode has been attributed to the
bond bending oxygen vibrations in the CuO$_{{\rm 2}}$ plane and,
when the orthorhombic distortion takes place, it splits to one
B$_{{\rm 2}{\rm u}}$ at $\sim $353 cm$^{{\rm -} {\rm 1}}$ and
another B$_{{\rm 3}{\rm u}}$ mode at 364 cm$^{{\rm -} {\rm 1}{\rm}
}$ \cite{Bazhenov}. Such splitting is not obvious in the Raman
spectra, but, as it is mentioned previously, the $\sim $370
cm$^{{\rm -} {\rm 1}}$ band is broad ($\sim $30 cm$^{{\rm -} {\rm
1}}$) (see Fig.9) and, also, seems to have some structure in the
low temperature spectra of the orthorhombic phase (Fig.2c). The
fitting of this band with two Lorentzians results in two peaks
that differ by $\sim $5-15 cm$^{{\rm -} {\rm 1}}$$^{{\rm} }$,
which agrees with the splitting of the two IR-active modes. Thus,
the $\sim $370 cm$^{{\rm -} {\rm 1}{\rm} }$ band could be another
IR active mode that appears in the xx/yy Raman spectra because of
the symmetry breaking. It should be noted that the largest
splitting ($\sim $15 cm$^{{\rm -} {\rm 1}}$) of the $\sim $370
cm$^{{\rm -} {\rm 1}{\rm} }$ band is observed for the underdoped
compound (x=0.03) at low temperatures, the main reason being the
large value of orthorhobicity, which makes the b axis much larger
than the a axis for this sample \cite{Radaelli,Lampakis}. The
energy of the B$_{{\rm 2}{\rm u}}$ vibrations of the Cu-O$_{xy}$
bond along the b axis is expected to be lower than the energy of
the B$_{{\rm 3}{\rm u}}$ vibrations along the a axis. Also, the
average softening of the total band at $\sim $370 cm$^{{\rm -}
{\rm 1}}$ with decreasing temperature that has been observed for
the x=0.03 sample (see Fig.9) could be due to a decrease of the
energy of the B$_{{\rm 2}{\rm u}}$ mode, since the difference
between the b and a axis increases with decreasing temperature
\cite{Radaelli}. The latter can also explain the increase of the
bandwidth for temperatures below $\sim $50 K compared with other
concentrations. If we assume the formation of stripes and the
folding of the Brillouin zone, an edge phonon that could be
related with the $\sim $370 cm$^{{\rm -} {\rm 1}}$ band would be
an X point mode of similar energy related with the apical oxygen
\cite{Pintschovius}.

The mode at $\sim $370 cm$^{{\rm -} {\rm 1}}$ gains intensity
together with the broad band at $\sim $280 cm$^{{\rm -} {\rm 1}}$,
and the wide mode at $\sim $125 cm$^{{\rm -} {\rm 1}}$, which
appear mainly at low temperatures (Fig.2). In previous IR studies
there is a weak B$_{{\rm 3}{\rm u}}$ phonon at $\sim $270
cm$^{{\rm -} {\rm 1}}$ \cite{Sugai,Bazhenov} attributed to the
vibrations of the Cu atoms in the ab-plane \cite{Bazhenov}. The
broad band at $\sim $280 cm$^{{\rm -} {\rm 1}}$ has not been
observed in the IR spectra of the doped La$_{{\rm 2}{\rm -} {\rm
x}}$Sr$_{{\rm x}}$CO$_{{\rm 4}}$ compounds
\cite{Collins,Bazhenov,Sugai}. In our compounds this band does not
appear in the Raman spectra of the underdoped x=0.03 concentration
even at very low temperatures (Fig.2a). It shows-up with
increasing Sr concentration and appears even in the high
temperature spectra of the tetragonal phase of the x=0.07, 0.092,
0.125, and 0.17 samples. For the overdoped compounds x=0.20 and
0.24, which are expected to be tetragonal at any temperature, this
band is absent to the lowest temperature studied (Fig.2e). Based
on its behavior, it is not possible to assign the very broad $\sim
$280 cm$^{{\rm - }{\rm 1}}$ band to a specific IR mode activated
in the Raman spectra from the breaking of symmetry. McQueeney et
al. based on inelastic neutron scattering of La$_{{\rm 2}{\rm -}
{\rm x}}$Sr$_{{\rm x}}$CuO$_{{\rm 4}}$ (0.0$ \le $0.15) at 10 K
(orthorhombic phase) have reported a systematic development of
anomalous phonon bands near the doping that induces metal to
insulator phase transition \cite{McQueeney}. From their phonon
density of states (PDOS) data, it can be seen that the
metal-insulator transition affects the low energy region, where a
band near 30 meV develops close in energy to the 280 cm$^{{\rm -
}{\rm 1}}$ band that appears in the Raman spectra. Also, the
intensity of $\sim $30 meV is maximized for optimally doped
samples, just like the 280 cm$^{{\rm -} {\rm 1}}$ mode. Inelastic
neutron scattering measurements showed that the $\sim $30 MeV band
is not caused by electrostatic impurity effects from Sr
substitution \cite{McQueeney2}. This band has not been observed in
any of the Raman measurements performed on samples with x=0-0.03
and becomes obvious for higher doping levels. Since the broad band
at $\sim$280 cm$^{{\rm -} {\rm 1}}$ always appears together with
the splitting of the $\sim$150 cm$^{{\rm -} {\rm 1}}$ and the wide
mode at $\sim$125 cm$^{{\rm -} {\rm 1}}$, and its energy fits well
with the combination energies of the two modes at $\sim$125
cm$^{{\rm -} {\rm 1}}$ and the $\sim$144 cm$^{{\rm -} {\rm 1}}$,
it is possible to be related with a peak in the PDOS. Its
appearance mainly at low temperatures whenever the multiple modes
around 150 cm$^{{\rm -} {\rm 1}}$ develop, could also indicate
that it might originate from quasiparticle scattering, which in
the Raman spectra would create such a high energy peak
\cite{stripes}.

Finally we examine the dependence of the relative intensity of the
$\sim $150, 370, and 280 cm$^{{\rm -} {\rm 1}}$ bands over the
intensity of the apex phonon at various temperatures. It is
already known that there is an apparent correlation of the
intensity of the $\sim $150 cm$^{{\rm -} {\rm 1}}$ peak with the
variation of the superconducting transition temperature T$_{{\rm
c}}$ with Sr content \cite{Lampakis}. Figure 13 shows the
variation of the above mode intensities at various temperatures
(concerning the $\sim $280 cm$^{{\rm -} {\rm 1}}$ band the room
temperature data are not included, since this peak has been
observed systematically only in the low temperature spectra). It
can be easily seen that the above mentioned correlation with
T$_{{\rm c}}$ persists even at the lowest temperature studied (10
K). \textit{From the above results, it can be seen that these
intensities mainly depend on the amount of doping and the local
lattice distortions induced by doping}. Another important effect
is the temperature where the broad band at $\sim$280 cm$^{{\rm -}
{\rm 1}}$ and the splitting of the band at $\sim$150 cm$^{{\rm -}
{\rm 1}}$ appear (Fig.2), which varies with doping; it vanishes
for x$<$0.05 or x$\geq$0.20 and shows a maximum around optimal
doping.

As it is already mentioned above, Bianconi et al, has proposed the
existence of charge stipes in the optimally doped LSCO, supporting
the 'two component' model, where, at optimum doping with 0.2
hole/Cu sites, a first component with hole density $\delta_{i}\sim
0.16$ coexists with another component of impurity states with
$\delta_{l} \sim 0.04$, spatially separated in two different types
of stripes forming a superlattice of quantum wires
\cite{Bianconi}. It is known that charge stripes break
translational (perpendicular to the stripes) and rotational
symmetry \cite{Kivelson}. The appearance of the above mentioned
modes in the Raman spectra seem to be induced by a symmetry
breaking.

On the other hand, it was suggested that the formation of
Jahn-Teller real space pairing can induce a local breaking of
symmetry and the appearance of symmetry-forbidden modes in the
Raman spectra \cite{Kabanov}. The formation of such extended
polaron complexes is related with stripes and it has a doping
dependence that resembles the variations observed, i.e. the
splitting of the $\sim $150 cm$^{{\rm -} {\rm 1}}$ band and the
appearance of the broad peak at $\sim $280 cm$^{{\rm -} {\rm 1}}$
in the doping range 0.20$>$x$>$0.03. It seems likely that below a
certain temperature the preformed polarons create a long range
order in one direction only and the modes split into a narrow and
a broad band \cite{stripes,Kabanov}. One cannot exclude also the
possibility that spin stripes are formed, which break the
spin-rotational and time reversal invariance \cite{Kivelson} and
coexist with a charge order \cite{Zachar}.

\subsection*{CONCLUSIONS}

In this work it is shown that the A$_{{\rm g}}$ mode at $\sim $100
cm$^{{\rm -} {\rm 1}}$ shows a classical soft mode behavior even
for high Sr-doped compounds (up to x=0.17), supporting its
correlation with the LTO$\rightarrow$HTT transition for the Sr
contents studied. Also, the mode at $\sim $270 cm$^{{\rm -} {\rm
1}{\rm} }$ of A$_{{\rm g}}$ symmetry due to vibrations of the
plane oxygens along the c axis correlates with the structural
phase transition since it appears only in the spectra of the
orthorhombic phase. Further, the energy of this mode varies with
temperature contrary to the results of previous works.

The low temperature measurements have shown that the behavior of
the energy and the asymmetry of the apex phonon should be due to
local lattice distortions, which are doping induced. The
modification of the relative intensity of the La/Sr and the apex
phonons seems to be correlated with the carrier concentration on
the CuO$_{{\rm 2}}$ planes and consequently with the
superconducting transition temperature.

The bands at $\sim $150 and 370 cm$^{{\rm -} {\rm 1}}$ that appear
in the xx/yy polarization spectra are related with a symmetry
breaking that occurs at temperatures that scale with T$_{c}$. The
first band splits in at least two modes from which, one is very
narrow and the other very wide and the spectra are reminiscent of
those obtained from the formation of stripes. Another broad band
was observed in the xx/yy polarization spectra at $\sim $280
cm$^{{\rm -} {\rm 1}}$, which appears to scale with the T$_{c}$
dependence on doping. This is probably due to a peak in the phonon
density of states or to quasiparticle scattering. Moreover, the
correlation of the relative intensity of the bands at $\sim $150
and 370 cm$^{{\rm -} {\rm 1}}$ with the Sr concentration for
maximum T$_{\textrm{c}}$ that was found at room temperature seems
to persist at the lowest temperature studied. A similar connection
exists between the intensity of the $\sim $280 cm$^{{\rm -} {\rm
1}}$ band and the doping. All these results can be attributed to
the formation of stripes at temperatures well above the transition
temperature and for doping levels where the compound is
superconducting.

\subsection*{ACKNOWLEDGEMENTS}
The team from NTUA expresses its appreciation to the Greek
Ministry of Education for financial support through the project
"Pythagoras I," co-funded by the European Social Fund ($75\%$) and
Greek National Resources ($25\%$). C.P. and the work in Cambridge
were supported by the Royal Society. We thank J. Cooper for
providing some of the samples studied. Critical remarks by D.
Mihailovic are also appreciated.

\bigskip

\subsection*{FIGURE CAPTIONS}

Figure1. Typical Raman spectra for selected temperatures for the
x=0.03 (a), x=0.092 (b), x=0.125 (c), x=0.17 (d), and x=0.24 (e)
samples in the y(zz)$\bar {y}$ scattering geometry using the 514.5
nm excitation wavelength.

\bigskip

Figure 2. Typical Raman spectra for selected temperatures for the
x=0.03 (a), x=0.092 (b), x=0.125 (c), x=0.17 (d), and x=0.24 (e)
samples in the y(xx)$\bar{y}$ (or x(zz)$\bar{x}$) scattering
geometry using the 514.5 nm excitation wavelength.

\bigskip

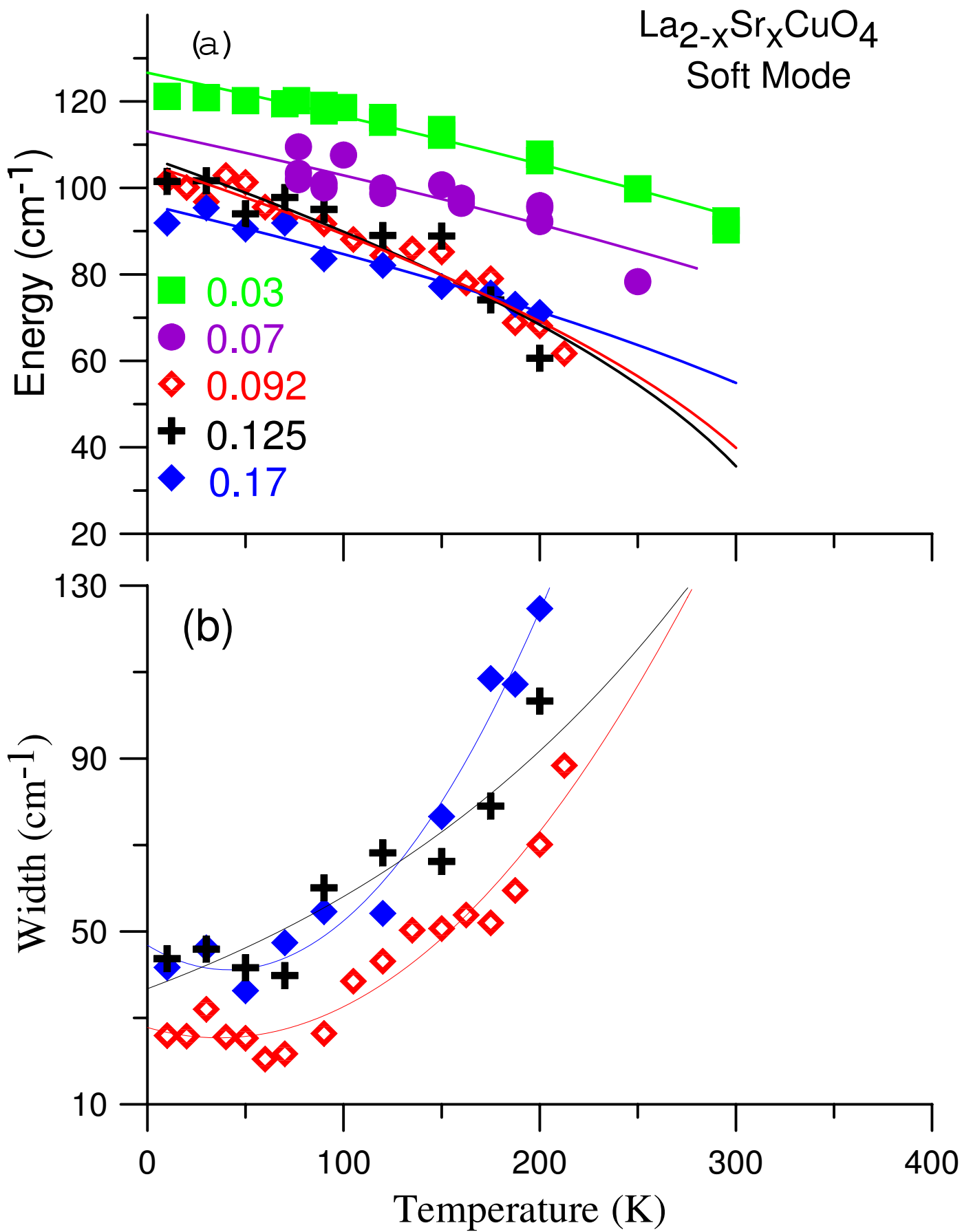
Figure 3. Dependence of the energy (a) and the linewidth (b) on
temperature for the soft phonon observed in y(zz)$\bar{y}$
scattering geometry spectra for x=0.03, 0.092, 0.125, 0.17, and
0.24.

\bigskip

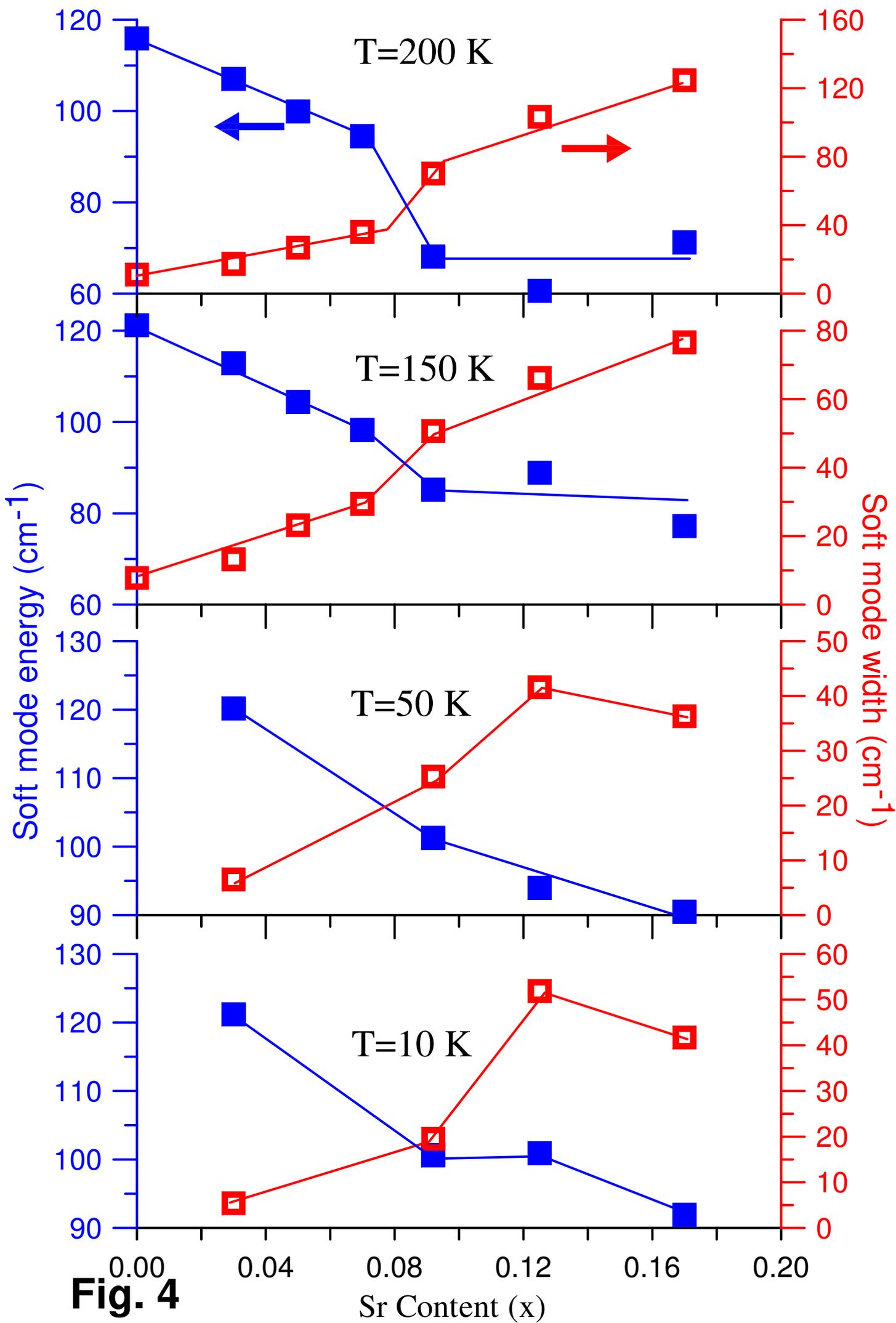
Figure 4. The dependence of the soft phonon (a) energy and (b)
linewidth on doping at various temperatures.

\bigskip

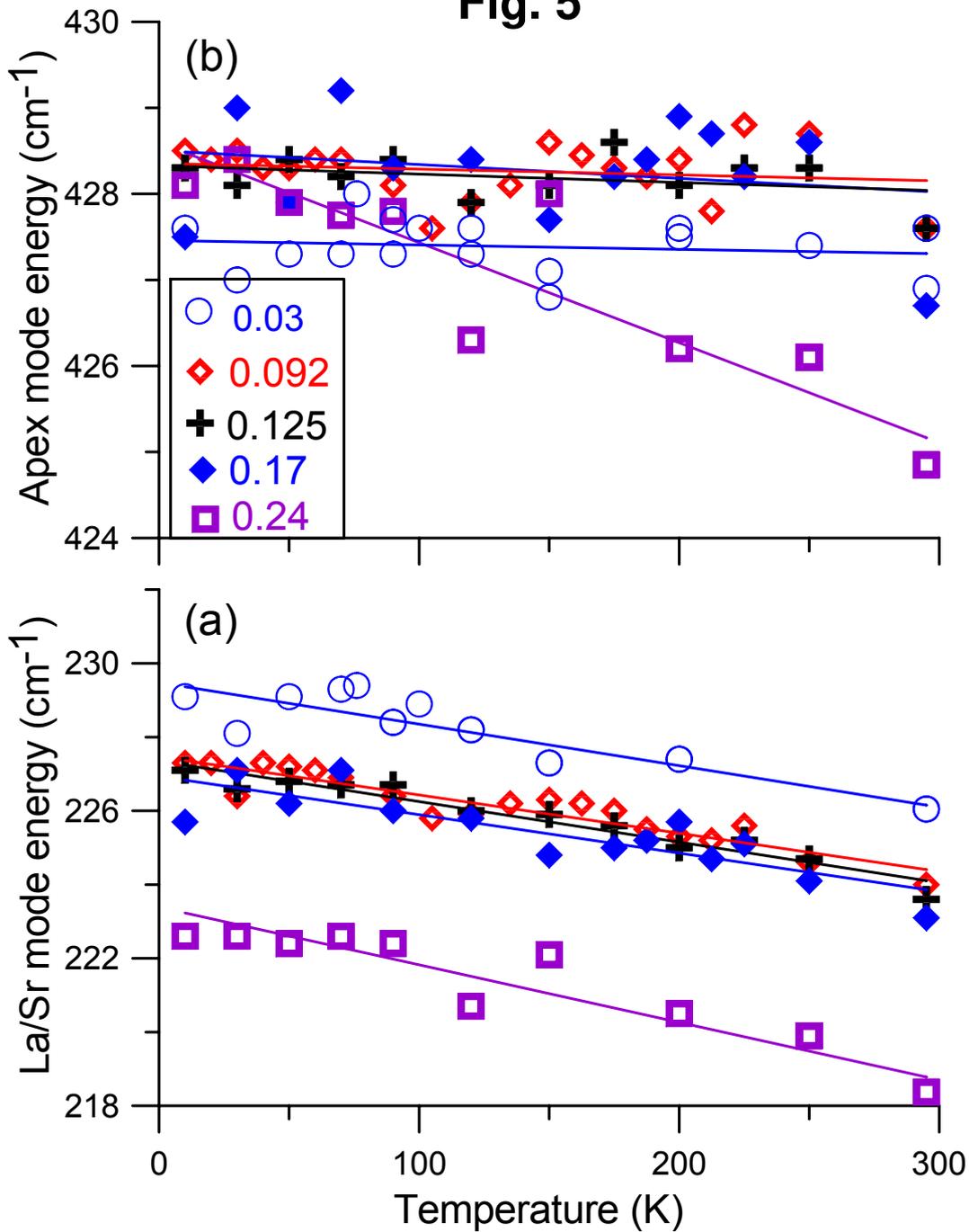
Figure 5. The dependence on temperature of the (a) La/Sr and (b)
apex phonon energy for the x=0.03, 0.092, 0,125, 0.17, and 0.24
samples.

\bigskip

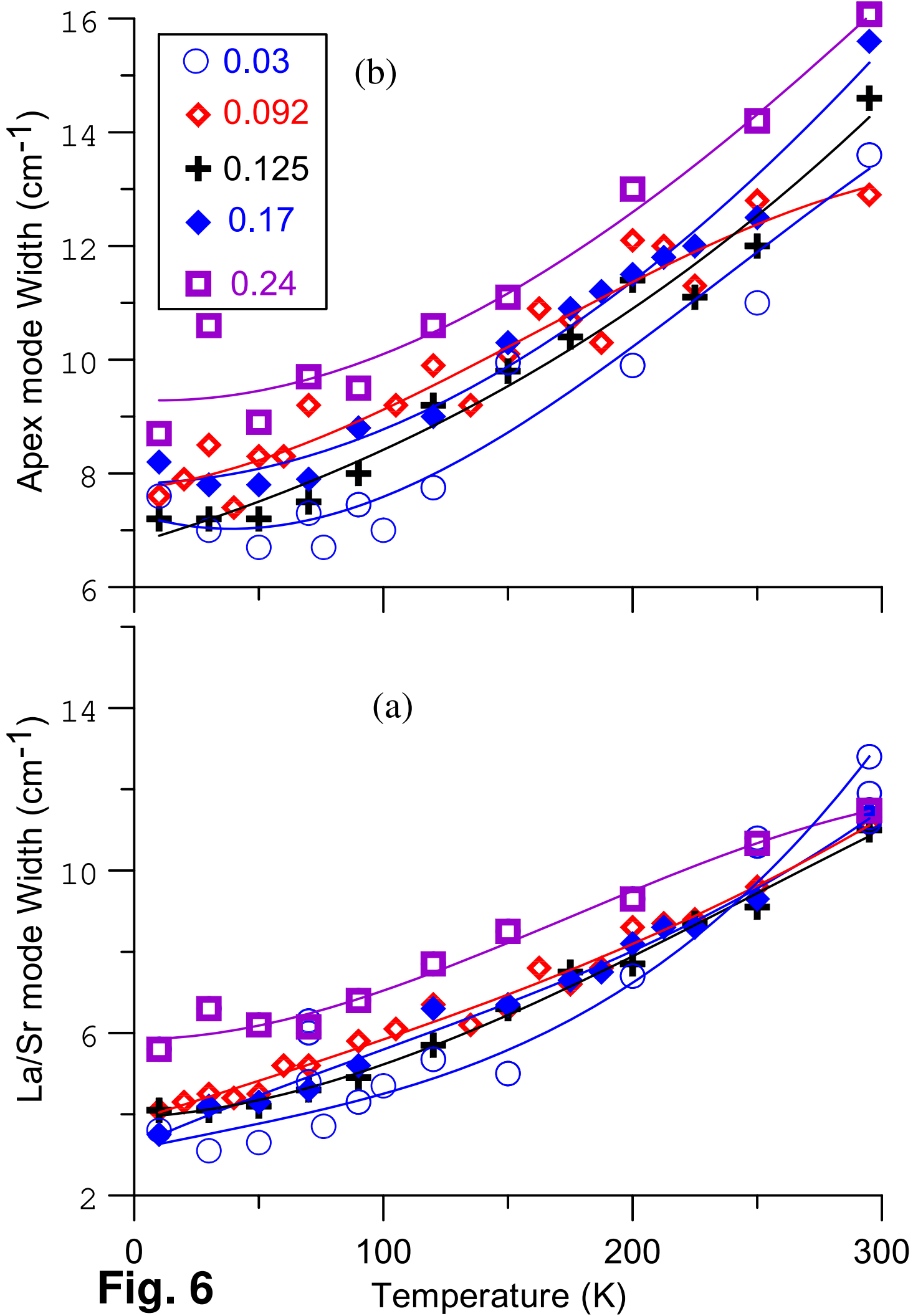
Figure 6. The dependence on temperature of the (a) La/Sr and (b)
apex phonon linewidth for the x=0.03, 0.092, 0,125, 0.17, and 0.24
samples.

\bigskip

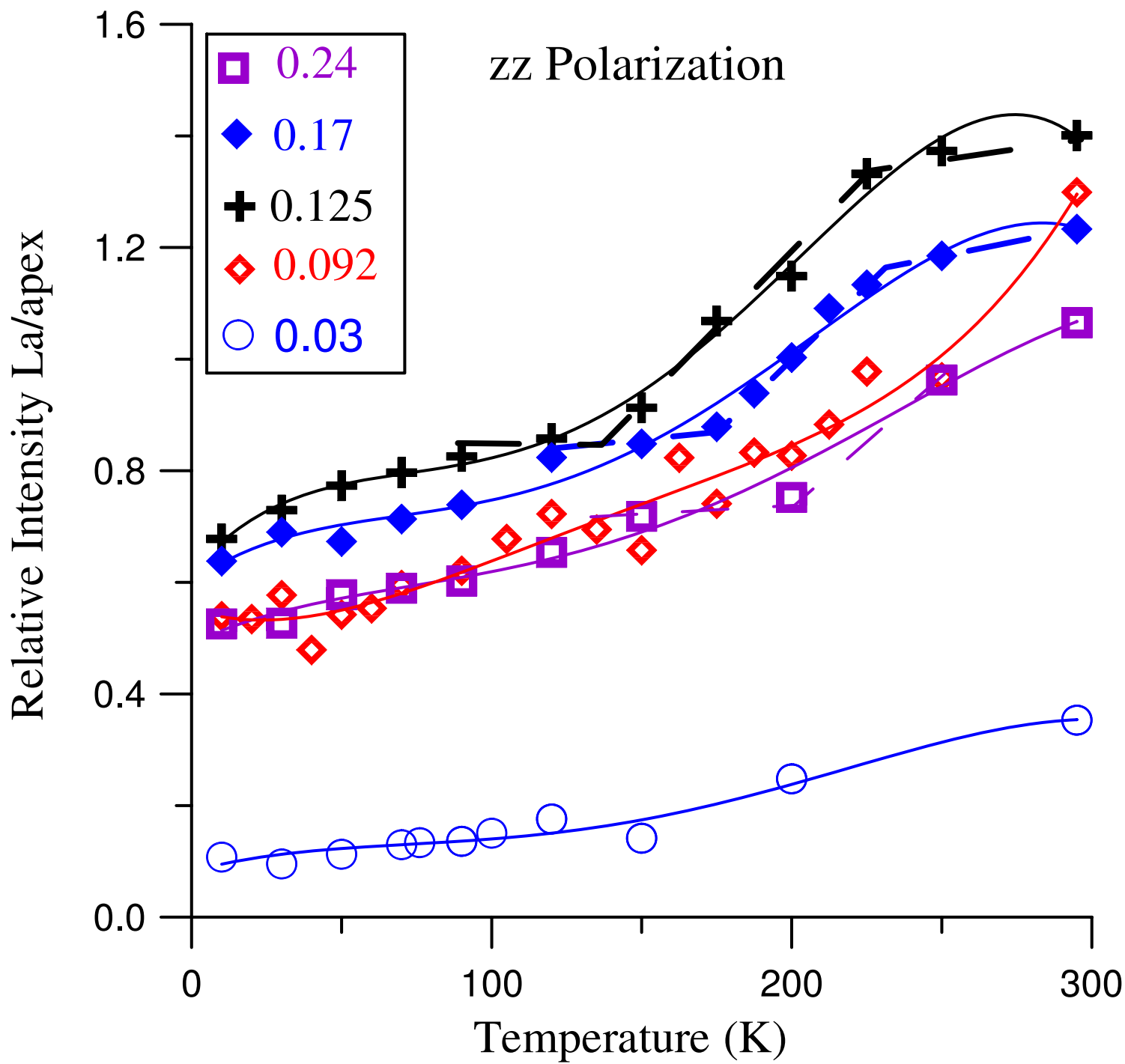
Figure 7. The relative intensity of the La/Sr phonon to the apex
one as a function of temperature for the x=0.03, 0.092, 0,125,
0.17, and 0.24 samples.

\bigskip

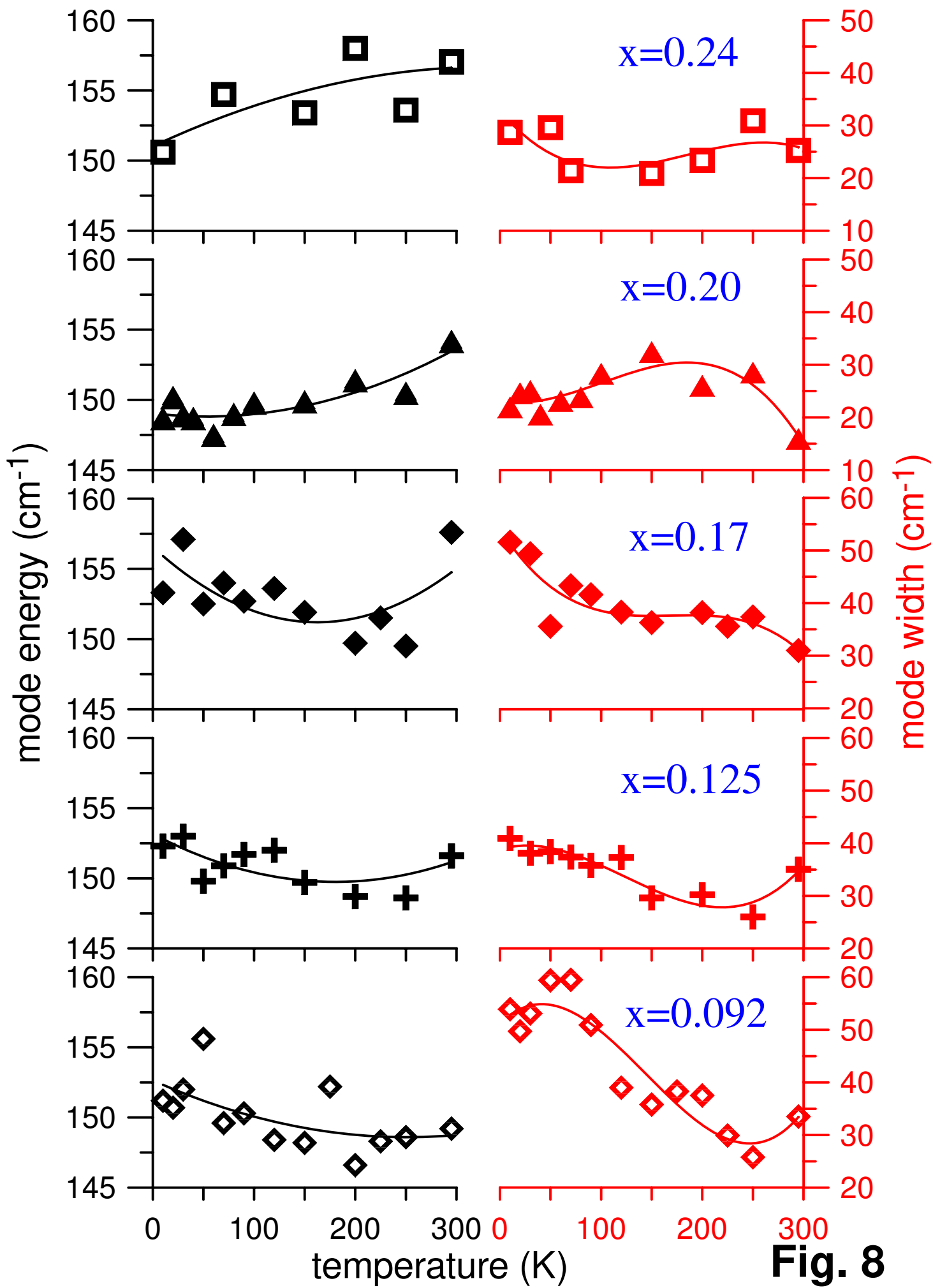
Figure 8. (a) Energy and (b) linewidth dependence on temperature
for the mode at $\sim $150 cm$^{{\rm -} {\rm 1}}$ observed in
y(xx)$\bar{y}$ scattering geometry spectra for x=0.092, 0.125,
0.17, 0.20, and 0.24.

\bigskip

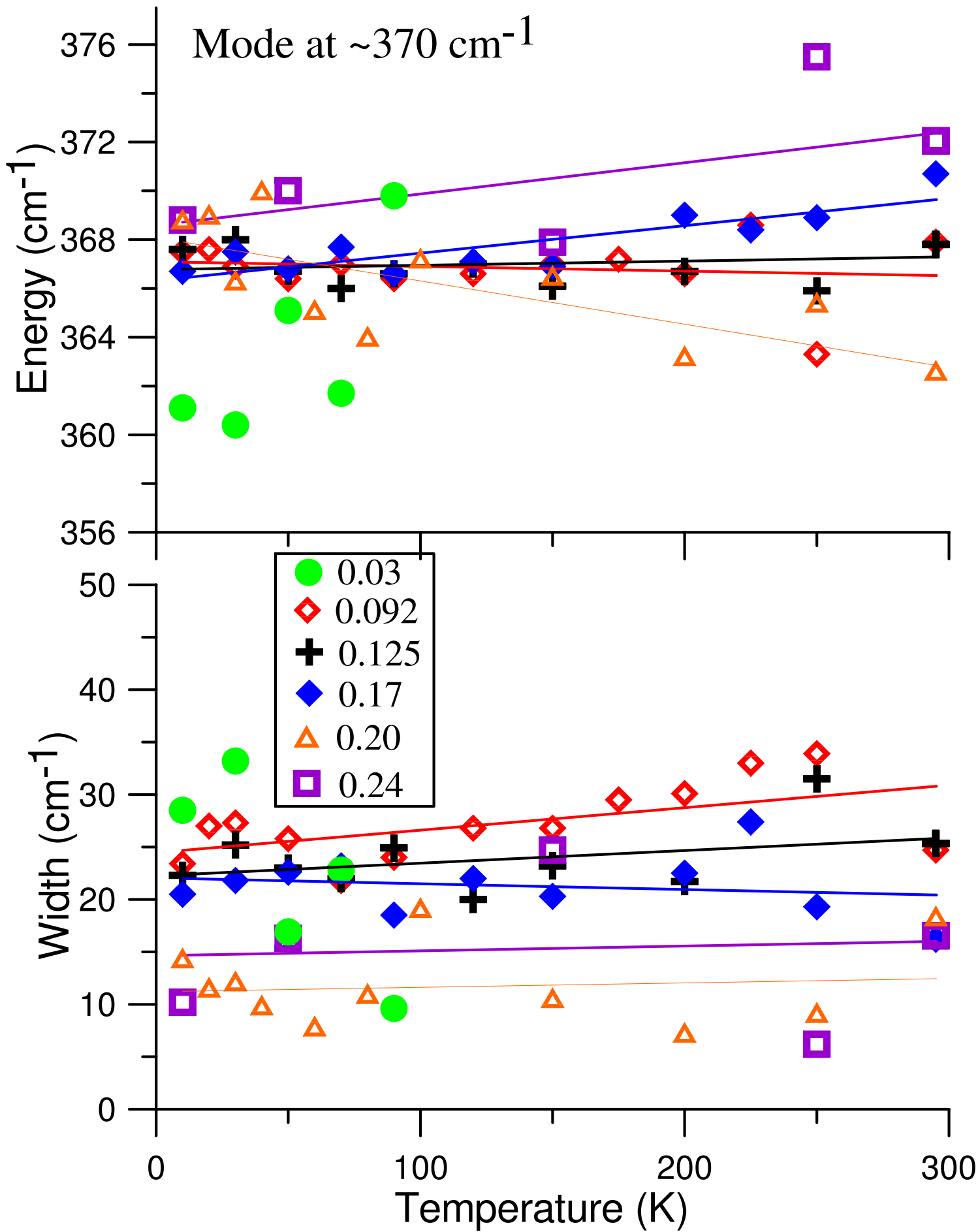
Figure 9. (a) Energy and (b) linewidth dependence on temperature
for the mode at $\sim $370 cm$^{{\rm -} {\rm 1}}$ observed in
y(xx)$\bar{y}$ scattering geometry spectra for x=0.03, 0.092,
0.125, 0.17, 0.20, and 0.24.

\bigskip

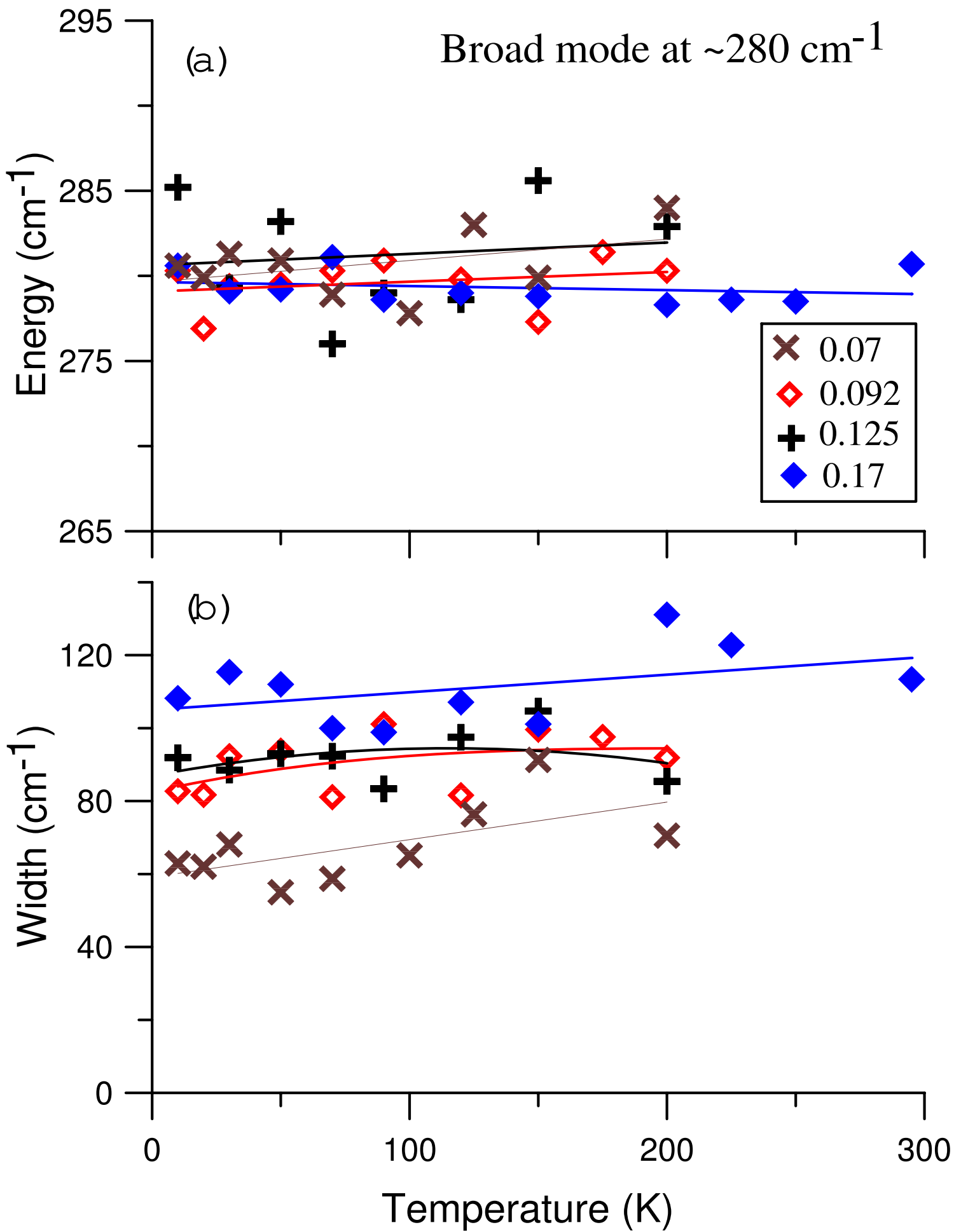
Figure 10. (a) Energy and (b) linewidth dependence on temperature
for the mode at $\sim $280 cm$^{{\rm -} {\rm 1}}$ observed in
y(xx)$\bar{y}$ scattering geometry spectra for x=0.03, 0.092,
0.125, 0.17, and 0.24.

\bigskip

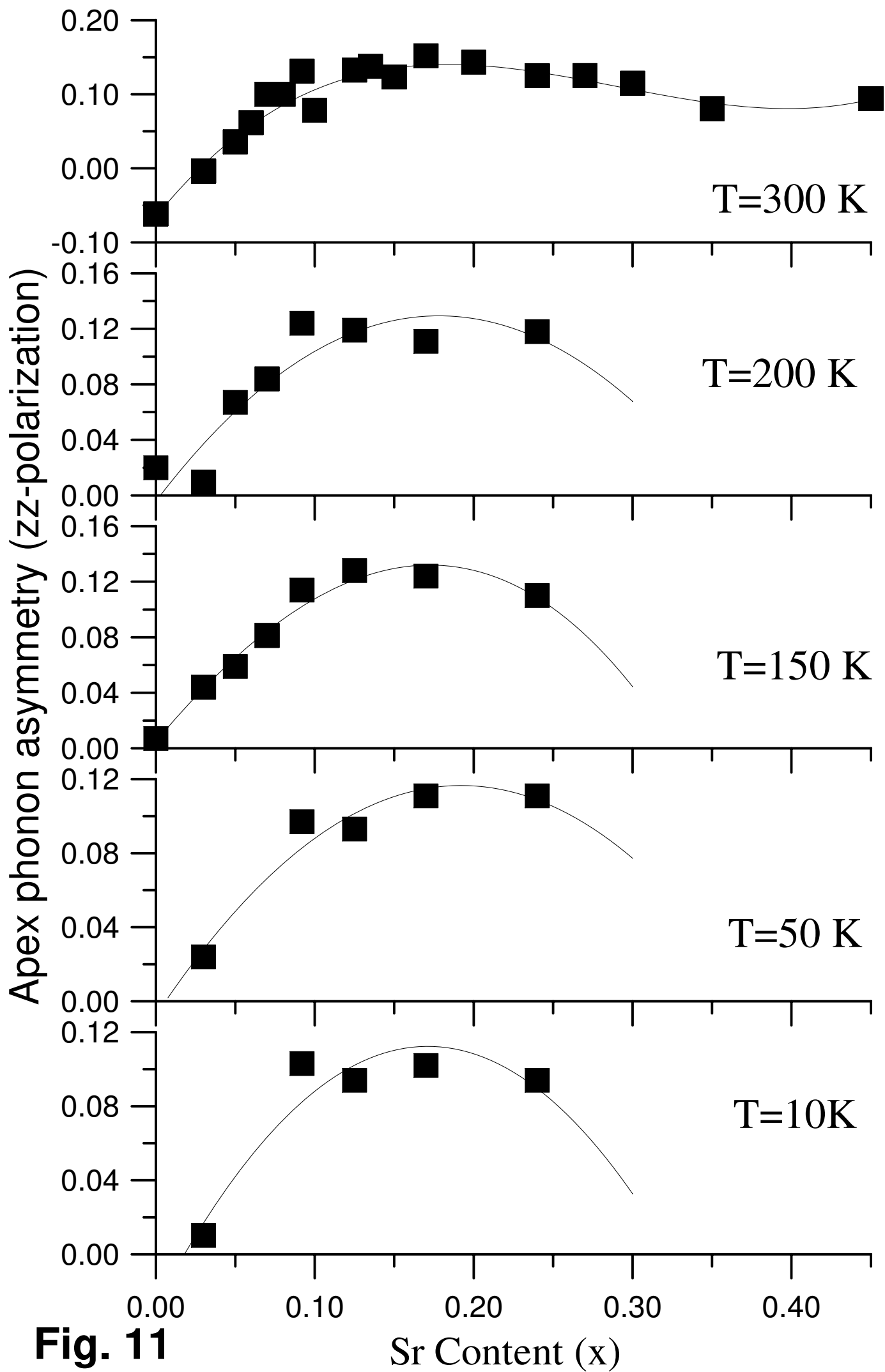
Figure 11. The dependence of the apex phonon asymmetry on doping,
at various temperatures.

\bigskip

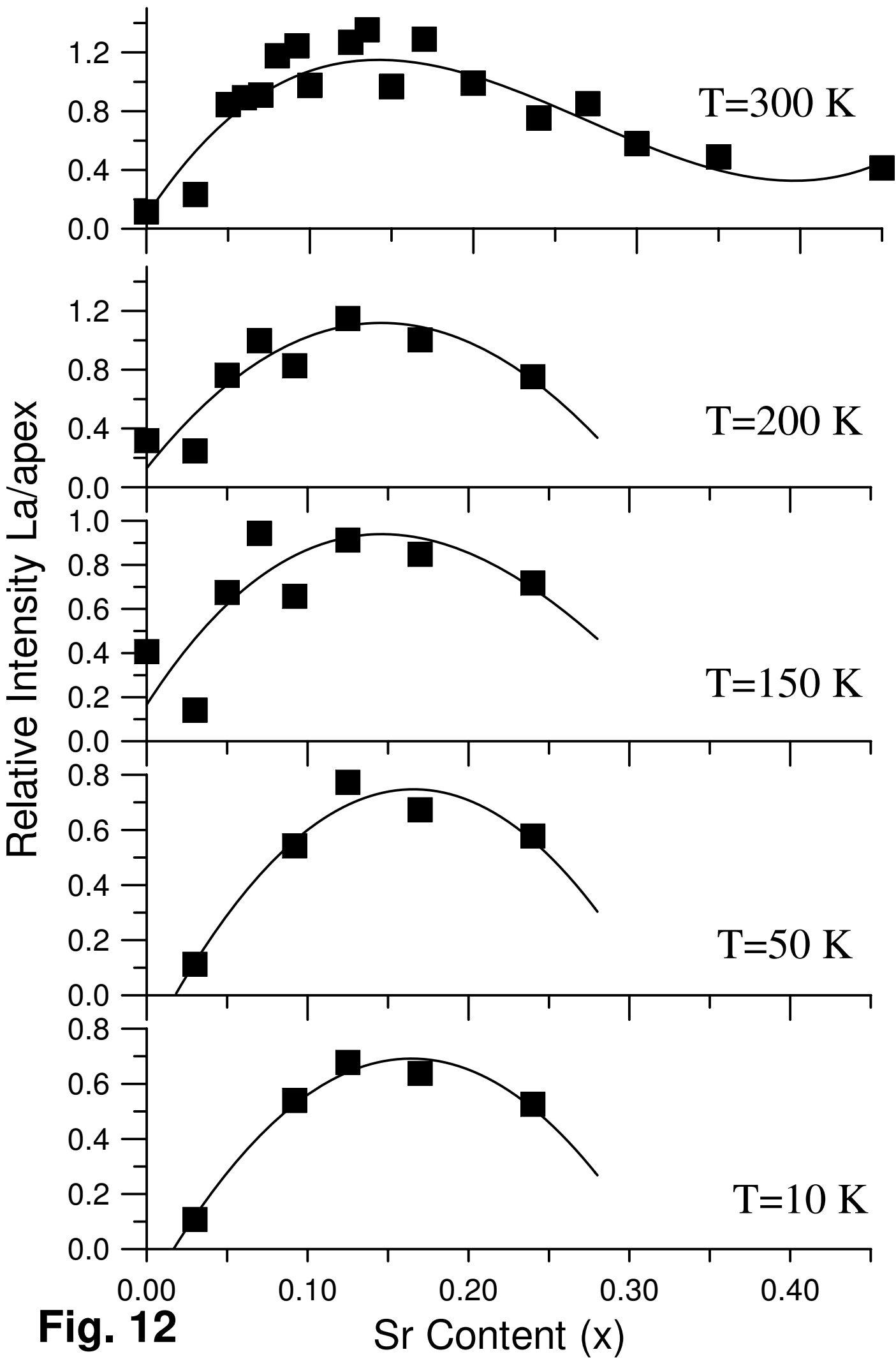
Figure 12. The dependence of the relative intensity La/apex, on
doping at various temperatures.

\bigskip

Figure 13. The dependence of the relative intensities (a)150
cm$^{{\rm -} {\rm 1}}$/apex, (b) 370 cm$^{{\rm -} {\rm 1}}$/apex,
and (c) 280 cm$^{{\rm -} {\rm 1}}$/apex on doping at various
temperatures.

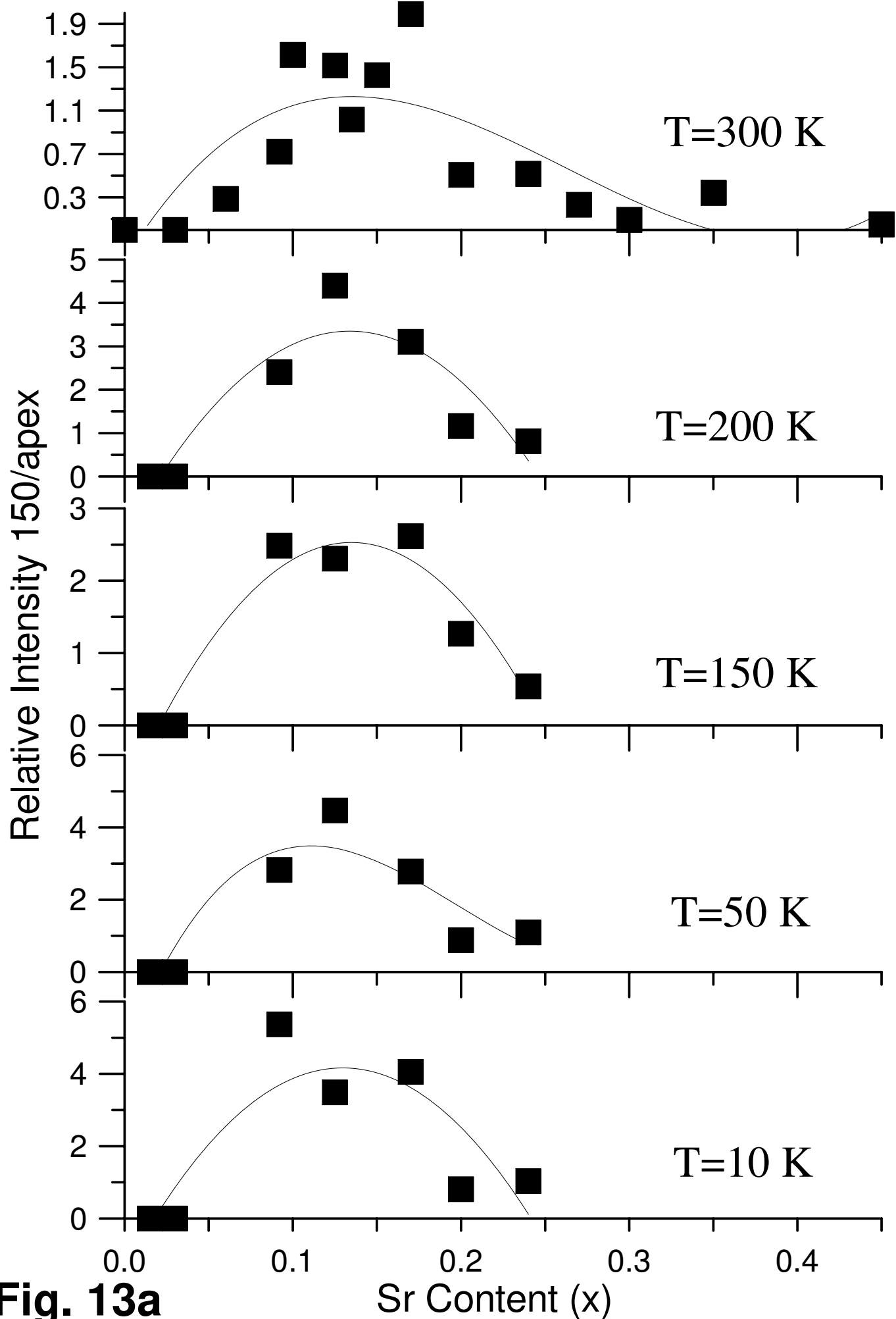
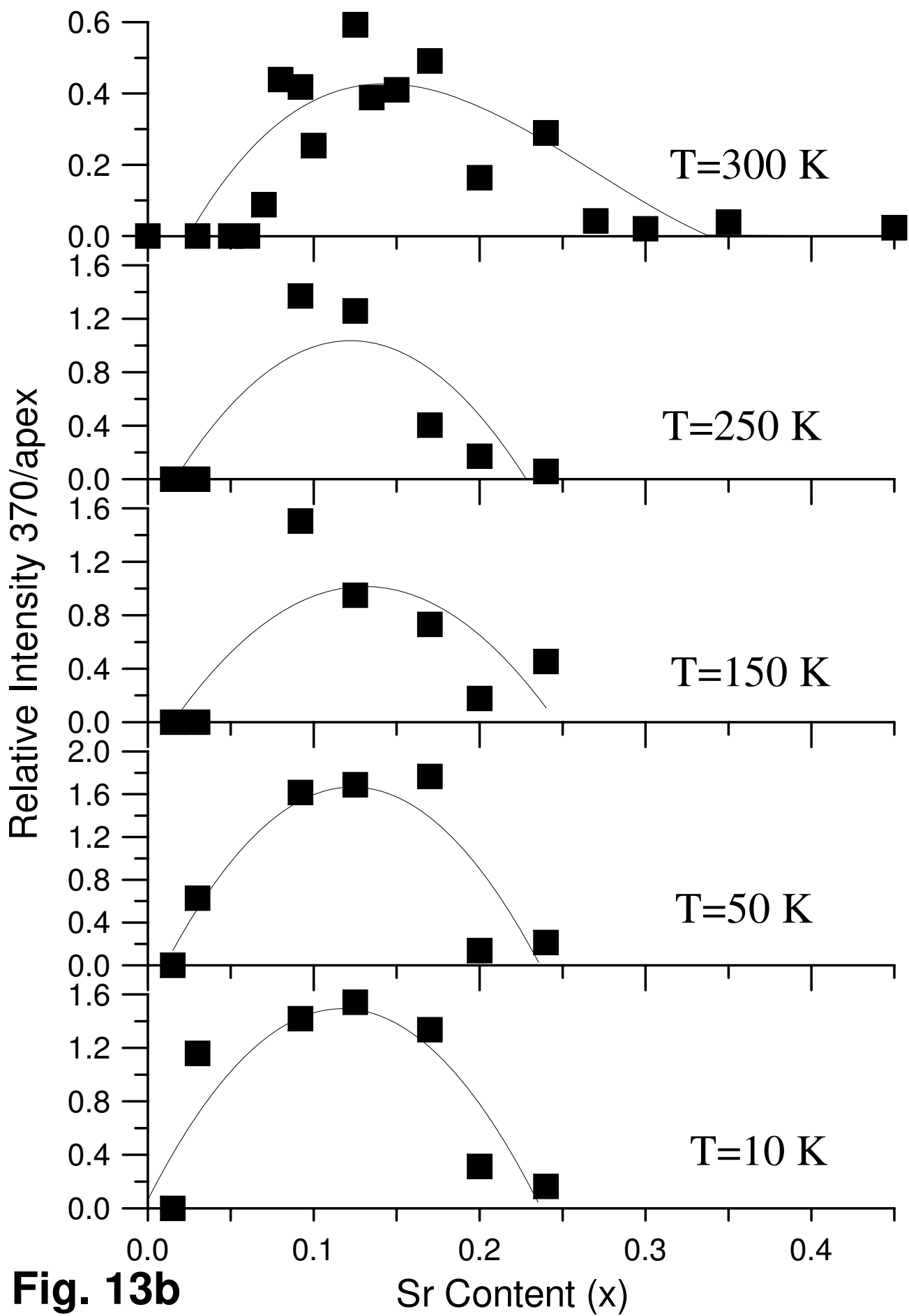
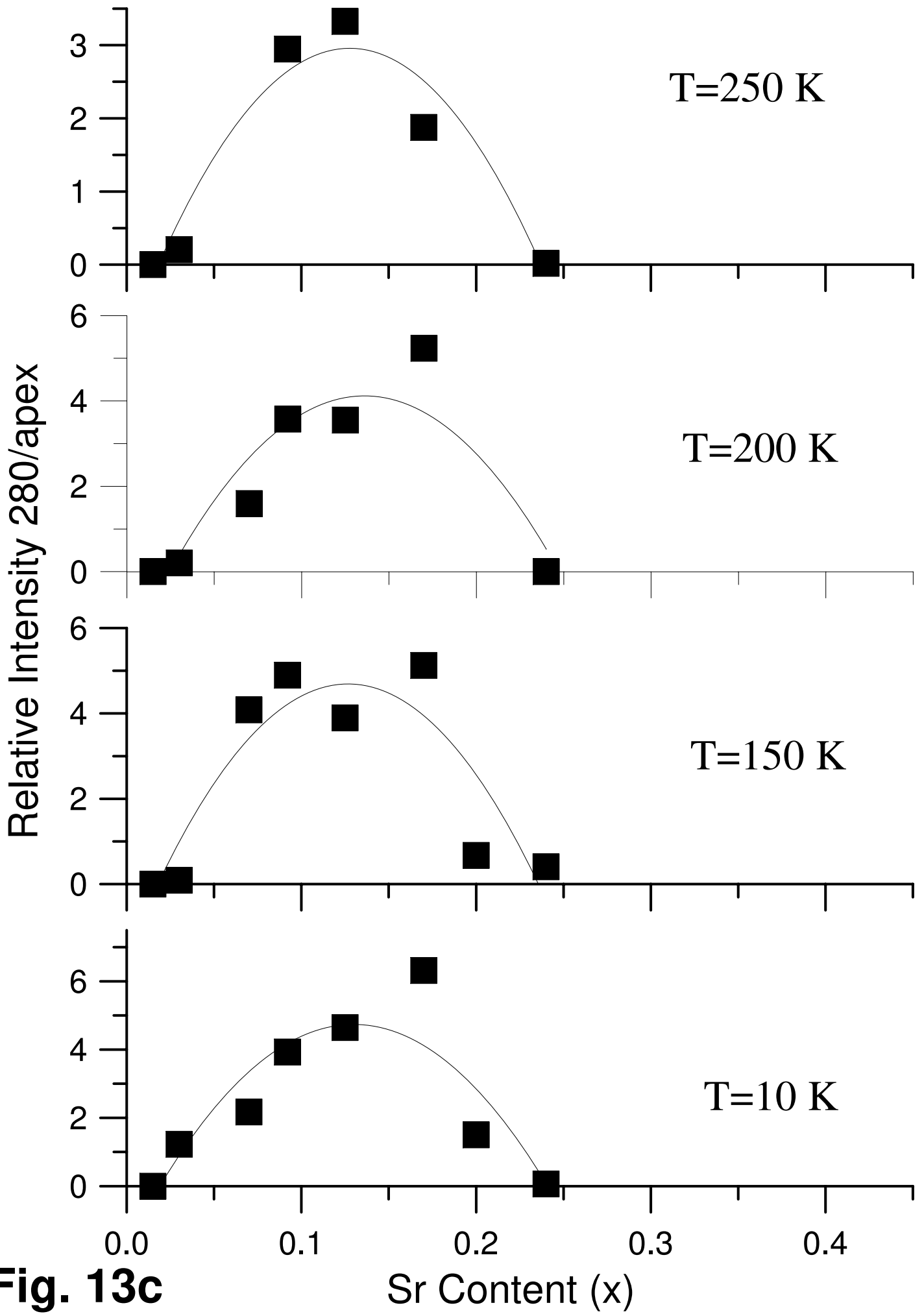
\end{document}